\documentclass{aastex}

\usepackage{amsmath,amssymb,amsthm,amscd,latexsym,w-greek}
\usepackage{hyperref,natbib}
\usepackage[toc,title,titletoc]{appendix}
\usepackage{cite}
\usepackage{graphicx,epsfig}
\RequirePackage{color}

\allowdisplaybreaks[1]

\def\Pb{P_{\rm b}}

\def\J2{J_2}

\def\virg#1{``#1"}

\def\eqi{\begin{equation}}
\def\eqf{\end{equation}}
\def\eqia{\begin{eqnarray}}
\def\eqfa{\end{eqnarray}}
\def\Om{\mathit{\Omega}}

\def\lb#1{\label{#1}}

\newcommand{\emaila}{lorenzo.iorio@libero.it}
\begin{document}
\title{Perspectives on effectively constraining the location of a massive trans-Plutonian object with the New Horizons spacecraft: a sensitivity analysis}
\shortauthors{L. Iorio}

\author{Lorenzo Iorio\altaffilmark{1} }
\affil{Ministero dell'Istruzione, dell'Universit\`{a} e della Ricerca (M.I.U.R.)-Istruzione\\
Fellow of the Royal Astronomical Society (F.R.A.S.)
\\ Permanent address for correspondence: Viale Unit\`{a} di Italia 68, 70125, Bari (BA), Italy.}

\email{\emaila}


\begin{abstract}
The radio tracking apparatus of the New Horizons spacecraft, currently traveling to the Pluto system where its arrival is scheduled for July 2015, should be able to reach an accuracy of 10 m (range) and 0.1 mm s$^{-1}$ (range-rate) over distances up to 50 au. This should allow to effectively constrain the location of a putative trans-Plutonian massive object, dubbed Planet X (PX) hereafter, whose existence has recently been postulated for a variety of reasons connected with, e.g., the architecture of the Kuiper belt and the cometary flux from the Oort cloud. Traditional scenarios involve a rock-ice planetoid with $m_{\rm X}\approx 0.7m_{\oplus}$ at some $100-200$ au, or a Jovian body with $m_{\rm X}\lesssim 5m_{\rm J}$ at about $10,000-20,000$ au; as a result of our preliminary sensitivity analysis, they should be detectable by New Horizons since they would impact its range at a km level or so over a time span six years long. Conversely, range residuals statistically compatible with zero having an amplitude of 10 m would imply that PX, if it exists, could not be located at less than about 4,500 au ($m_{\rm X}=0.7m_{\oplus}$) or 60,000 au ($m_{\rm X}=5m_{\rm J}$), thus making a direct detection quite demanding with the present-day technologies. As a consequence, it would be appropriate to rename  such a remote body as \textit{Telisto}. Also fundamental physics would benefit from this analysis since certain subtle effects predicted by MOND for the deep Newtonian regions of our Solar System are just equivalent to those of a distant pointlike mass.
\end{abstract}




\section{Introduction}
The possible existence of a distant, pointlike object having the mass of a planet, or even  more, in the remote peripheries of our Solar System is traditionally well rooted in the scientific community for a number of different reasons which are summarized in, e.g., \citep{2012CeMDA.112..117I}.  Several names were proposed so far for such a putative object according to its hypothesized physical nature and role, ranging from a rock-ice planet with mass comparable to that of Mars and the Earth to a brown/red dwarf: Planet X \citep{Lowell1915}, Nemesis \citep{1984PNAS...81..801R,2010MNRAS.407L..99M}, Tyche \citep{2011Icar..211..926M}. For the sake of simplicity, we will conventionally denote it as Planet X (PX hereafter); after our analysis, it will turn out that it may  quite appropriately be renamed as\footnote{From $\uptau\acute{\upeta}\uplambda\upiota\upsigma\uptau \upo\upvarsigma:$  \textit{farthest, most remote.}} \textit{Telisto}. It is interesting to note that, in recent years,  the PX hypothesis is back in vogue \citep{2006Icar..184..589G,2008AJ....135.1161L,2011ApJ...726...33F,2012DDA....43.0501G,2012arXiv1212.6124S}. As an example, Lykawka  \citep{2012arXiv1212.6124S} pointed out that the fine orbital structure of the Kuiper belt, the orbits of both the Jovian and the Neptunian Trojans, and, perhaps, the current orbits of the giant planets could be accommodated by  the orbital evolution of Mars/Earth-like primordial embryos. They were ultimately scattered by the giant planets; nonetheless, the presently observed features of the trans-Neptunian objects at 40-50 au are optimally satisfied if at least one such primordial planetoid survived in the outskirts of the Solar System.

On the other hand, the PX scenario is certainly not infrequent in several extrasolar systems, thus yielding new, stronger motivations about the fascinating possibility that the same may occur  in our Solar System as well. For example, the M dwarf GJ 317  hosts two Jupiter-size planets with semimajor axes $a_{\rm b}=1.148$ au and $a_{\rm c} = 10-40$ au, respectively \citep{2012ApJ...746...37A}. The multiple system of the young late A star HR 8799 is made of five huge Jovian planets ($7-10 m_{\rm J}$) orbiting at large astrocentric distances ($15-68$ au) \citep{0004-637X-741-1-55}. HIP 5158c is a $m=15m_{\rm J}$ planet moving at $7.7$ au from its parent star which is more closely ($a_{\rm b}=0.89$ au) orbited by another lighter Jupiter-type planet ($m= 1.44 m_{\rm J}$) \citep{2011MNRAS.416L.104F}. Single sub-stellar distant companions have been discovered as well; as an example, CT Cha ($M_{\star}=M_{\odot}$) has a brown dwarf-type partner ($m=17 m_{\rm J}$) at 440 au \citep{2008A&A...491..311S}. Even more distant is the brown dwarf candidate ($m=8m_{\rm J}$) WD 0806-661Bb, at 2500 au from its hosting star \citep{2011ApJ...732L..29R}. Moving to compact parents, PSR B1620-26 \citep{psr1,psr2} is a triple system made of a pulsar-white dwarf binary in a relatively close orbit ($\Pb= 102$ d) \citep{1996ASPC..105..525A} orbited by quite
distant circumbinary planet-like companion8 ($\Pb \sim 10^4$ d). In general, theoretical studies encounter difficulties in explaining the formation and origin of such planets at wide orbits ($>$100 au) around their host stars. Recently, it has been proposed \citep{2012ApJ...750...83P} that they can be the result of dynamical recaptures of free floating planets in dispersing stellar clusters and stellar associations. The discovery of a remote  trans-Plutonian planet would, thus, also contribute to shed light on the environment in which the formation of our Solar System took place \citep{2011MNRAS.417.2104V,2012MNRAS.421.2969V,2012MNRAS.422.1648V}.

PX investigations have also direct relevance  for fundamental physics and alternative theories of gravity; indeed, a distant, isolated mass such as a hypothetical planet would have the same signature as certain subtle effects predicted by MOND in the Solar System \citep{2009MNRAS.399..474M,2011MNRAS.412.2530B}. Moreover, PX can be viewed as a potentially non-negligible source of systematic bias on several proposed high accuracy tests of general relativity and fundamental physics to be conducted in the far regions of our Solar System. Analogous considerations hold for extrasolar binaries made of compact objects orbited by distant PX-type companions as well \citep{Freire012}.
Finally, PX studies have also connections with the issue of the three-dimensional Galactic acceleration of our Solar System \citep{2012A&A...544A.135X} which, in turn, is crucial for the (non-baryonic) Dark Matter problem.

In this paper, we focus on the constraints on the spatial location of PX which can indirectly be inferred from its putative gravitational pull on known
Solar System's objects. An analysis \citep{2012CeMDA.112..117I} using the supplementary precessions of the perihelia \citep{2010IAUS..261..159F} of some known planets of the Solar System was recently made to put constraints on the minimum allowable heliocentric distance of PX for different hypothesized values of its mass as a function of its ecliptic longitude and latitude. Although more recent data on extra-rates of planetary perihelia \citep{2011CeMDA.111..363F} appeared after the study in  \citep{2012CeMDA.112..117I}, here we will follow a different approach. We will investigate the future perspectives to enhance the present-day bounds on the location of PX offered by accurate radio tracking of anthropogenic objects approaching the known peripheries of the planetary regions of our Solar System. Indeed,  several spacecraft-based missions with such targets have recently been proposed and/or approved: Uranus Pathfinder \citep{2012ExA....33..753A}, OSS (Outer Solar System) \citep{2012ExA....34..203C}, New Horizons \citep{2008SSRv..140....3S}.

Among them, we will concentrate on New Horizons since it has been implemented.
It was launched on 19 January 2006, and it is currently \textit{en route} to   the Pluto system, where it should arrive in July 2015, and the Kuiper belt. Contrary to the Voyager 1-2 spacecrafts, New Horizons is spin-stabilized; thus, accurate radioscience will be possible \citep{lrr-2010-4}.
Thanks to REX \citep{2008SSRv..140..217T}, an uplink radio science instrument with radiometer capabilities, the accuracy in range tracking should be better than 10 m (1$\sigma$) up to 50 au \citep{2008SSRv..140...23F} corresponding to an extended mission configuration up to six years more after 2015, while Doppler velocity measurements accurate to better than $0.1$ mm s$^{-1}$ should be possible throughout the mission  \citep{2008SSRv..140...23F}.
This opens interesting perspectives to effectively constraining  the location of a putative remote PX through its gravitational effect on the  range $\rho$ and the range-rate $\dot\rho$ of New Horizons.
\section{PX-induced Range signatures}
Here, we  will produce  numerically integrated signals $\Delta\rho,\Delta\dot\rho$ induced by  PX on the range and range-rate of New Horizons for different values of its putative mass and spatial location. As far as the position of PX is concerned, we will parameterize it in terms of the usual Keplerian orbital elements which are the semimajor axis $a$, the eccentricity $e$, the inclination $I$ of the orbital plane to the reference $\{x,y\}$ plane adopted,  the longitude of the ascending node $\Om$,  the argument of pericenter $\omega$, and the true anomaly $f$. More specifically,  we will first numerically integrate the equations of motion in rectangular Cartesian coordinates of the Earth, PX and New Horizons in a heliocentric coordinate system having the mean ecliptic at J2000.0 ad reference $\{x,y\}$ plane; the initial conditions, corresponding to, say, 1 Jan. 2015, are retrieved from the HORIZONS WEB interface maintained by NASA. We will also compute the Earth-New Horizons range and the range-rate by direct differentiation. Then, we  will repeat the same procedure without PX in the model, and we will take the difference of the two numerically produced time series corresponding to the same initial conditions both for the range and the range-rate.
The time interval of the integration is 6 yr corresponding to an extended mission--to be approved by NASA--through the Kuiper Belt up to 50 au \citep{2008SSRv..140...23F}. The resulting signals $\Delta\rho,\Delta\dot\rho$ are representative of the impact of PX on the New Horizons directly observable quantities.

We will use such templates to preliminarily infer an order-of-magnitude evaluation of the constraints on PX which could plausibly be obtained from the  range and range-rate residuals of New Horizons. Clearly, it is just a preliminary sensitivity analysis which in no ways pretends to replace an accurate covariance analysis based on simulations of the actual data reduction procedure which, among other things, should require an explicit inclusion  of PX in the dynamical models to be fit to the simulated observations.

As a naive rule-of-thumb, we will assume that, for a given value of the mass $m_{\rm X}$ of PX and of its spatial orientation, the minimum distance allowed for PX will be the one providing a peak-to-peak amplitude of the simulated signal as large as 10 m for the range and 0.1 mm s$^{-1}$ for the range-rate. If the peak-to-peak  amplitude of, say, the range signal will be  larger than 10 m, then the minimum distance of PX can actually be even larger, and vice-versa.

We will consider just two \virg{classical} scenarios for PX; the one involving a $0.7 m_{\oplus}$ rock-ice planetoid at some $100-175$ au \citep{2008AJ....135.1161L,2012arXiv1212.6124S}, and a Tyche-type giant planet ($1m_{\rm J}\lesssim m_{\rm X}\lesssim 5m_{\rm J}$) at about $10,000-30,000$ au \citep{2011Icar..211..926M,2011ApJ...726...33F}. In the following, we will only consider the range since it turned out that it yields tighter constraints than the range-rate.
\subsection{The Mars/Earth-type case}
It turns out that if a PX with $m_{\rm X}=0.7 m_{\oplus}$ is really at $100-175$ au, then it will hardly escape from detection by New Horizons since its simulated peak-to-peak range amplitude would be as large as $1.5-7$ km; perhaps, even a direct, raw inspection of the actual range residuals produced without even modeling its action would be able to detect it. Indeed, its signature would be so huge that it is unlikely that it could be removed from the real range residuals below the 10 m level due to a partial \virg{absorption} of its unmodeled signal in the fit of the initial conditions. However, such an issue will certainly deserve dedicated, more specific studies.

What could be concluded about PX if nothing different from zero at a statistically significative level will appear in the range residuals of New Horizons?
In Figure \ref{figura1}-Figure \ref{figura5} we depict the PX-induced New Horizons range signals over a time span six years long, corresponding to a heliocentric distance of New Horizons up to 50 au, for $m_{\rm X}=0.7 m_{\oplus}$ whose peak-to-peak amplitudes are as large as about 10 m by varying the parameters fixing its position in space.

It is significant to note that all the plots were obtained for  $a_{\rm X} = 4,500$ au. Even taking into account  the aforementioned caveat, this means that a null result would imply that an Earth-like rock-ice body could be located at not less than about $4,000$ au or, for some particular spatial orientation of the orbit of PX, even more (see Figure \ref{figura4}).
\subsection{The Tyche-type case}
The same kind of numerical analysis previously made for a rock-ice, Earth-sized planetoid shows that a Jovian mass at 10,000 au would impact the range of New Horizons with a signal having a peak-to-peak amplitude as large as 1 km over six years, while for $m_{\rm X}=5m_{\rm J}$ and $a_{\rm X}=20,000$ au the range perturbation would amount to $\approx 1.2$ km over the same time span. Thus, it can reasonably be concluded that also such a kind of PX would be detectable by New Horizons.

Also in this case, it is interesting to consider the consequences of a null result. From Figure \ref{figura6}-Figure \ref{figura10} it turns out that a brown dwarf object with $m_{\rm X}=5m_{\rm J}$ should be as distant as  $\approx 60,000$ au, or even more for certain orbital geometries,  to yield range residuals of the order of $\approx 10$ m or so.

We mention the fact that a Sun-like main sequence star should be located at $3\times 10^6$ au to affect the New Horizons range at a $\approx 10$ m level.

It is important to remark that if the range residuals of New Horizons will be statistically compatible with zero, this will have severe consequences about the possibility of directly detecting PX from its emitted electromagnetic radiation, either in the visible or in the infrared. Indeed, from the sensitivity analysis presented here and from the discussion on the direct detectability summarized in Section 4 of \citep{2012CeMDA.112..117I}, it turns out that imaging so remote bodies would be quite difficult with the present-day technologies.
\section{Summary and conclusions}
We preliminarily investigated the effects that a distant, pointlike massive object lurking in the outskirts of our Solar System, not yet directly discovered, would have on the range and range-rate of the New Horizons spacecraft over a time span of 6 yr in a possible extended mission configuration after its arrival at the Pluto system scheduled in July 2015. We assumed an accuracy of 10 m and $0.1$ mm s$^{-1}$ for the range and the range-rate of New Horizons, respectively.

It turned out that the range should provide tighter constraints on the location of such a putative trans-Plutonian object than the range-rate.

An Earth-sized rock-ice planetoid at some $\approx 100-200$ au or a Jovian mass at $\approx 10,000-20,000$ au should be relatively easy to detect since they would impact the range of New Horizons at a km level or so. Even if the dynamical action of such a body was not explicitly modeled in the actual data reduction process, it is unlikely that so huge signatures may be removed from the residuals below the 10 m level due to a possible partial signal \virg{absorption} in estimating the initial conditions. Conversely, a null result, i.e. range residuals statistically compatible with zero as large as 10 m, would have the consequence of greatly increasing the minimum allowable distance for a putative distant companion of the Sun. A $0.7m_{\oplus}$ planetoid should be located at approximately $4,500$ au, or even more depending on its orbital geometry, while a brown-dwarf with $m_{\rm X}=5m_{\rm J}$ could only exist at not less than about $60,000$ au. As a consequence, its direct detection would become quite demanding with the currently available technologies. As anticipated in the Introduction, naming \textit{Telisto} such a so far object would, thus, be quite appropriate.

The results of such an investigation would also be relevant for fundamental physics and modified models of gravity. Indeed, certain consequences  predicted by MOND for the External Field Effect in the deep Newtonian regions of our Solar System would just mimic the action of a distant, pointlike mass.

Finally, as a concluding remark, we will stress once again that the present study is a sensitivity analysis. A useful, complementary approach, which is beyond the scopes of this work, consists of a full covariance analysis involving the estimation of one or more dedicated parameters for \textit{Telisto} in the reduction of simulated data of New Horizons.
\renewcommand\appendix{\par
\setcounter{section}{0}%
\setcounter{subsection}{0}%
\setcounter{table}{0}
\setcounter{figure}{0}
\setcounter{equation}{0}
\gdef\thetable{\Alph{table}}
\gdef\thefigure{\Alph{figure}}
\gdef\theequation{\Alph{section}.\arabic{equation}}
\section*{Appendix}
\gdef\thesection{\Alph{section}}
\setcounter{section}{0}}

\appendix
\section{Figures}\lb{figurine}
In this Appendix, all the figures quoted in the text are collected for a better overall readability of the paper. Figure \ref{figura1}-Figure \ref{figura5} refer to the range signatures $\Delta\rho$ due to a planetoid with $m_{\rm X}=0.7\ m_{\oplus}$, while Figure \ref{figura6}-Figure \ref{figura10} deal with the case $m_{\rm X}=5\ m_{\rm J}$. They cover a 6-yr interval starting from $t_0 =$ 1 Jan 2015.
\begin{figure}
\centering
\begin{tabular}{ccc}
\epsfig{file=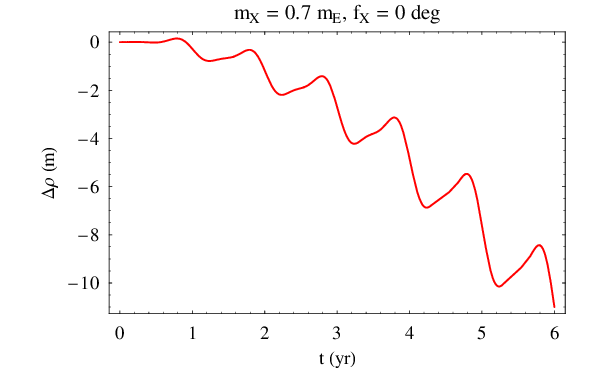,width=0.30\linewidth,clip=} & \epsfig{file=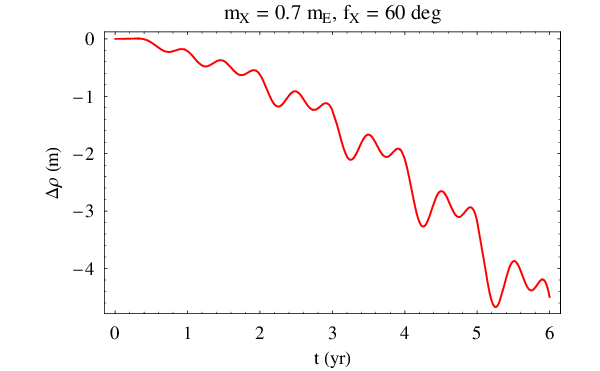,width=0.30\linewidth,clip=} &
\epsfig{file=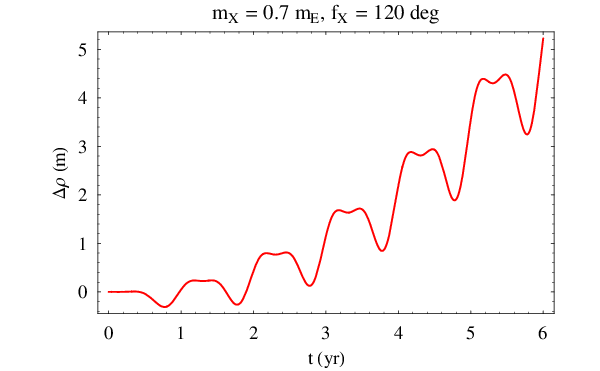,width=0.30\linewidth,clip=} \\
\epsfig{file=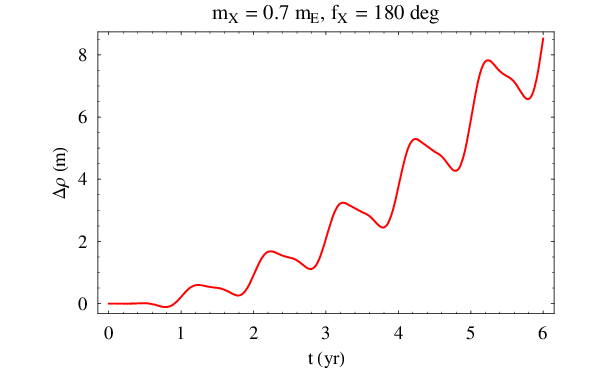,width=0.30\linewidth,clip=}&
\epsfig{file=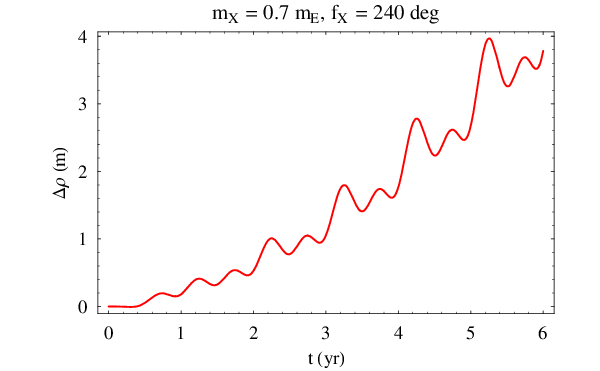,width=0.30\linewidth,clip=} & \epsfig{file=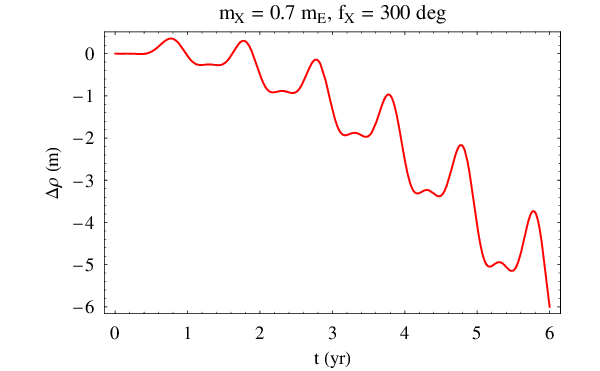,width=0.30\linewidth,clip=}\\
\end{tabular}
\caption{Numerically integrated Earth-New Horizons range signal $\Delta\rho$, in m, for $t_0= 1$ Jan 2015, $m_{\rm X}=0.7\ m_{\oplus}$, $a_{\rm X}=4,500$ au, $e_{\rm X}=0.07$, $I_{\rm X}=35$ deg, $\Om_{\rm X}=60$ deg, $\omega_{\rm X}=50$ deg.}\lb{figura1}
\end{figure}
\begin{figure}
\centering
\begin{tabular}{ccc}
\epsfig{file=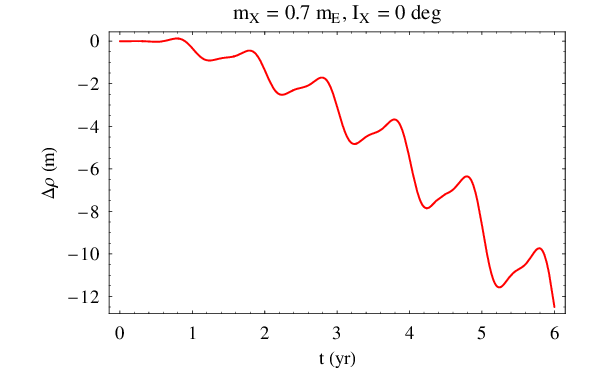,width=0.30\linewidth,clip=} & \epsfig{file=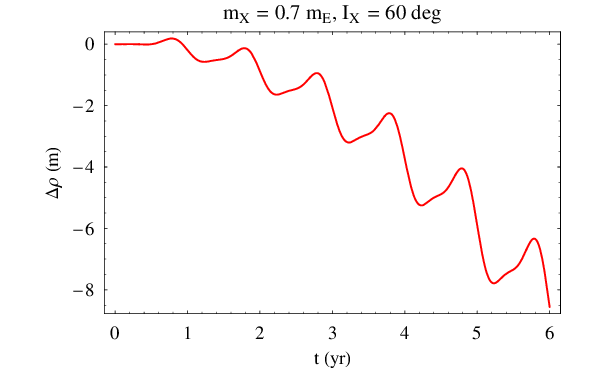,width=0.30\linewidth,clip=}&
\epsfig{file=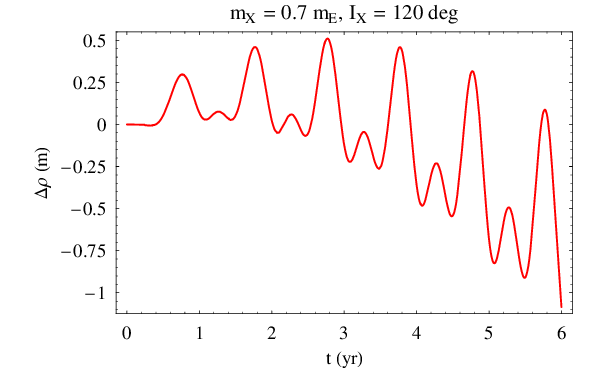,width=0.30\linewidth,clip=} \\
\epsfig{file=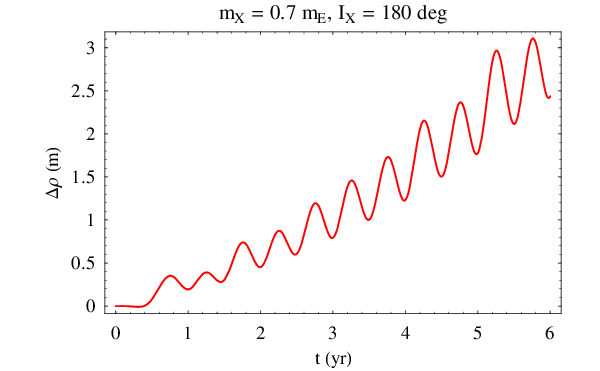,width=0.30\linewidth,clip=}&
\epsfig{file=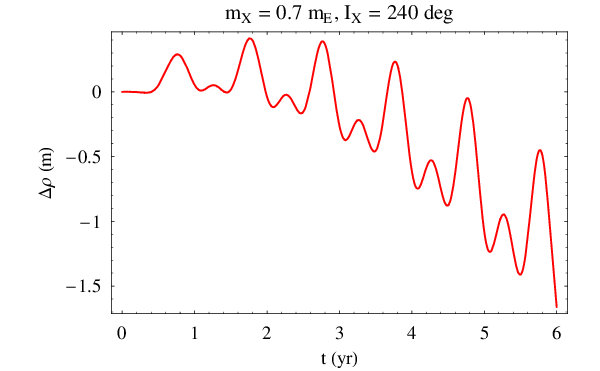,width=0.30\linewidth,clip=} & \epsfig{file=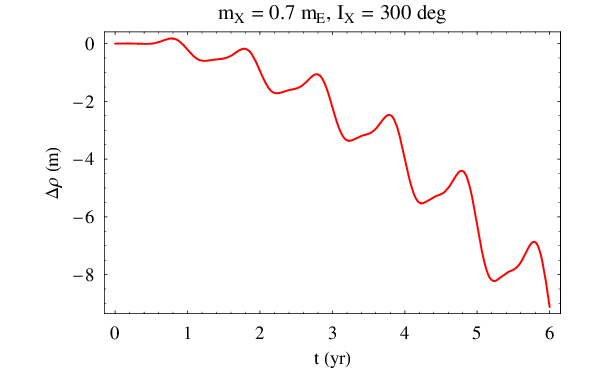,width=0.30\linewidth,clip=}\\
\end{tabular}
\caption{Numerically integrated Earth-New Horizons range signal $\Delta\rho$, in m, for $t_0= 1$ Jan 2015, $m_{\rm X}=0.7\ m_{\oplus}$, $a_{\rm X}=4,500$ au, $e_{\rm X}=0.07$, $f_{\rm X}=0$ deg, $\Om_{\rm X}=60$ deg, $\omega_{\rm X}=50$ deg.}\lb{figura2}
\end{figure}
\begin{figure}
\centering
\begin{tabular}{ccc}
\epsfig{file=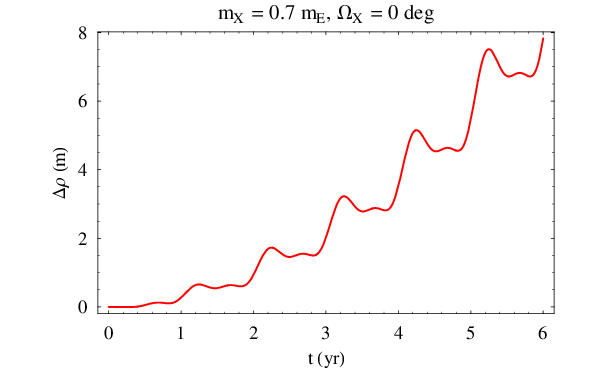,width=0.30\linewidth,clip=} & \epsfig{file=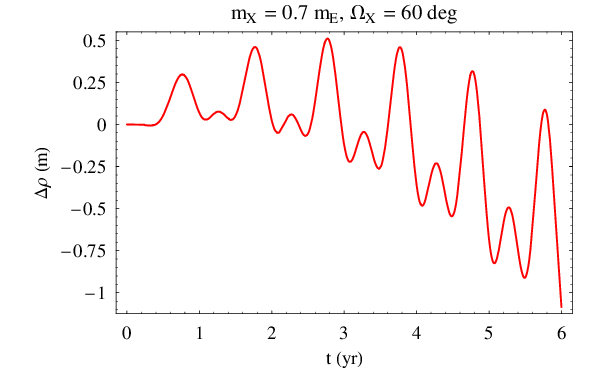,width=0.30\linewidth,clip=}&
\epsfig{file=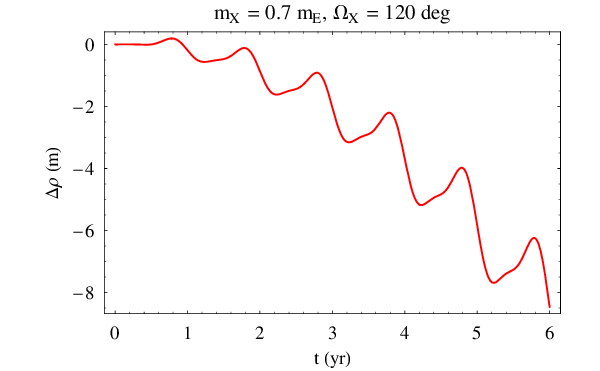,width=0.30\linewidth,clip=} \\
\epsfig{file=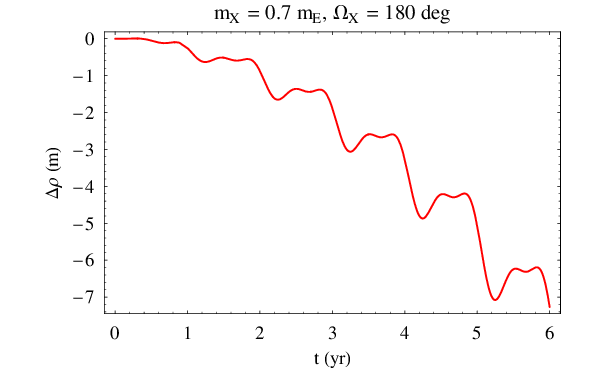,width=0.30\linewidth,clip=}&
\epsfig{file=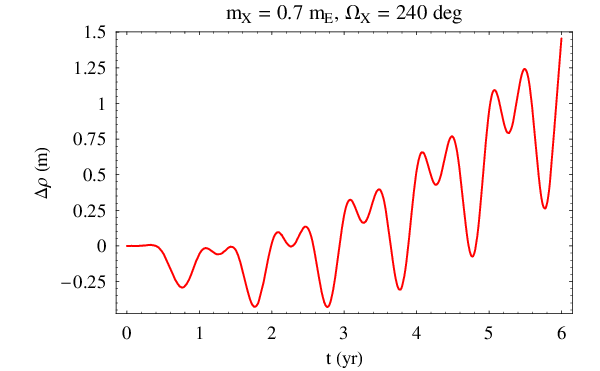,width=0.30\linewidth,clip=} & \epsfig{file=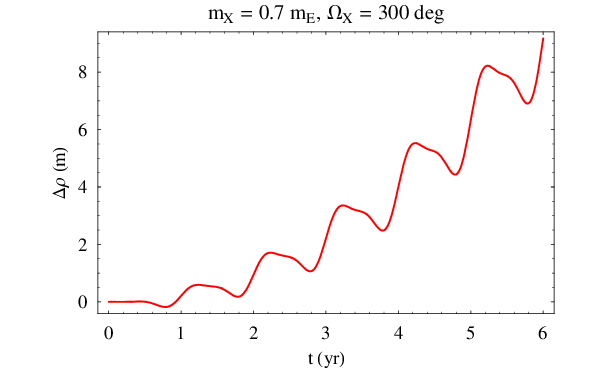,width=0.30\linewidth,clip=}\\
\end{tabular}
\caption{Numerically integrated Earth-New Horizons range signal $\Delta\rho$, in m, for $t_0= 1$ Jan 2015, $m_{\rm X}=0.7\ m_{\oplus}$, $a_{\rm X}=4,500$ au, $e_{\rm X}=0.07$, $f_{\rm X}=0$ deg, $I_{\rm X}=120$ deg, $\omega_{\rm X}=50$ deg.}\lb{figura3}
\end{figure}
\begin{figure}
\centering
\begin{tabular}{ccc}
\epsfig{file=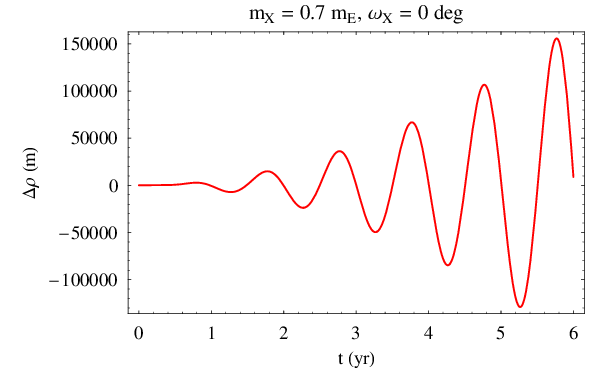,width=0.30\linewidth,clip=} & \epsfig{file=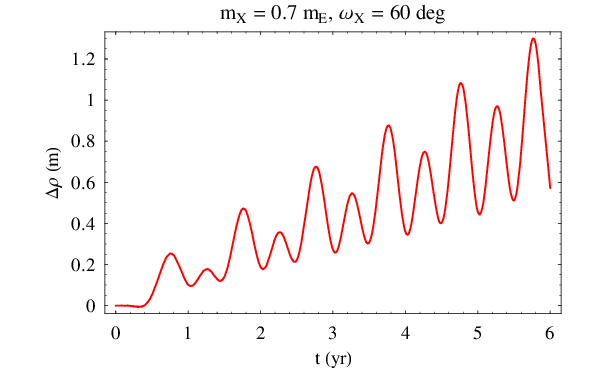,width=0.30\linewidth,clip=}&
\epsfig{file=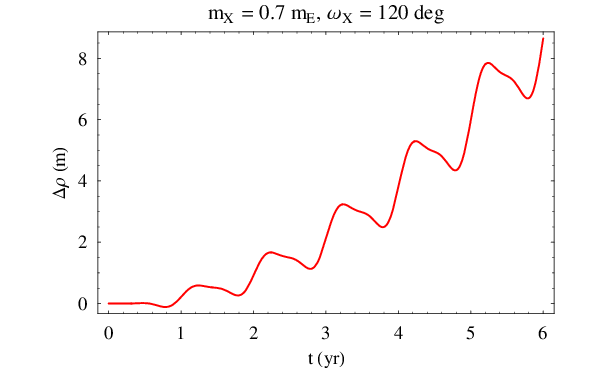,width=0.30\linewidth,clip=} \\
 \epsfig{file=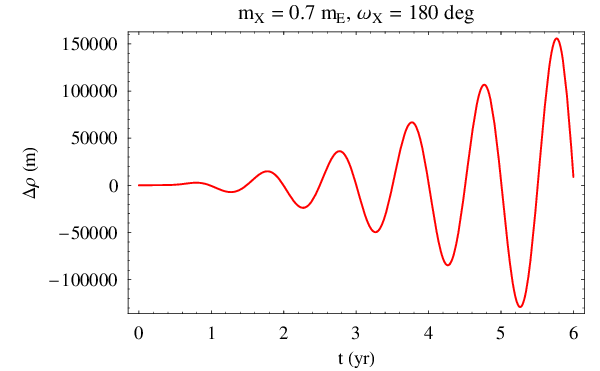,width=0.30\linewidth,clip=} &
\epsfig{file=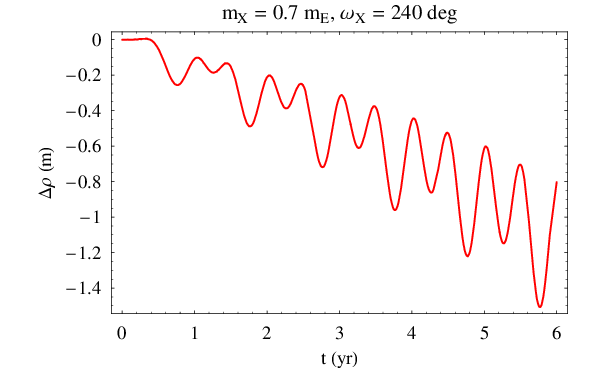,width=0.30\linewidth,clip=} & \epsfig{file=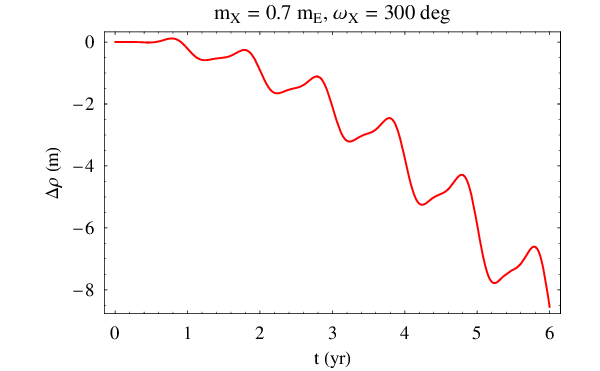,width=0.30\linewidth,clip=}\\
\end{tabular}
\caption{Numerically integrated Earth-New Horizons range signal $\Delta\rho$, in m, for $t_0= 1$ Jan 2015, $m_{\rm X}=0.7\ m_{\oplus}$, $a_{\rm X}=4,500$ au, $e_{\rm X}=0.07$, $f_{\rm X}=0$ deg, $I_{\rm X}=120$ deg, $\Om_{\rm X}=60$ deg.}\lb{figura4}
\end{figure}
\begin{figure}
\centering
\begin{tabular}{ccc}
\epsfig{file=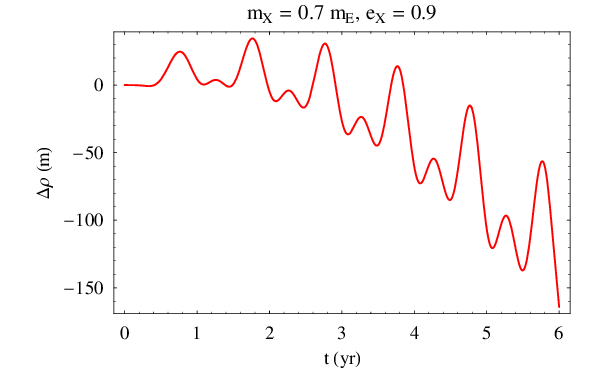,width=0.30\linewidth,clip=} & \epsfig{file=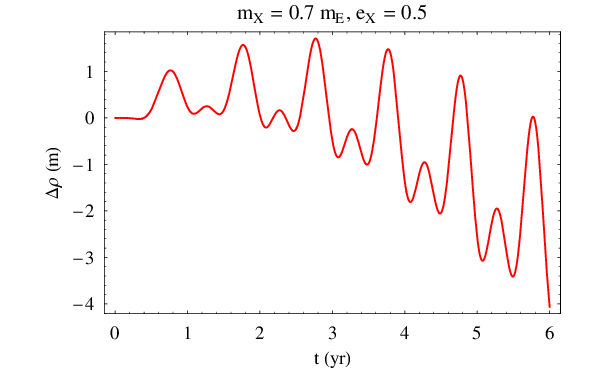,width=0.30\linewidth,clip=}&
\epsfig{file=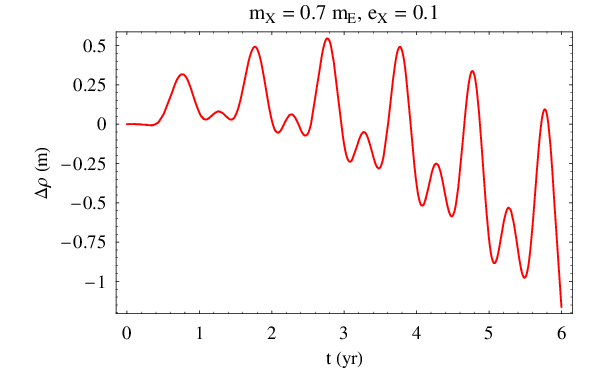,width=0.30\linewidth,clip=} \\
 \epsfig{file=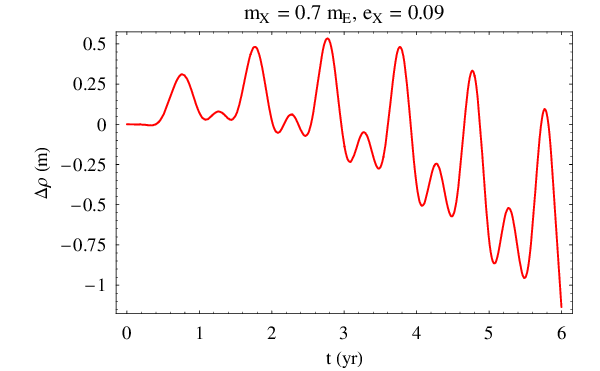,width=0.30\linewidth,clip=}&
\epsfig{file=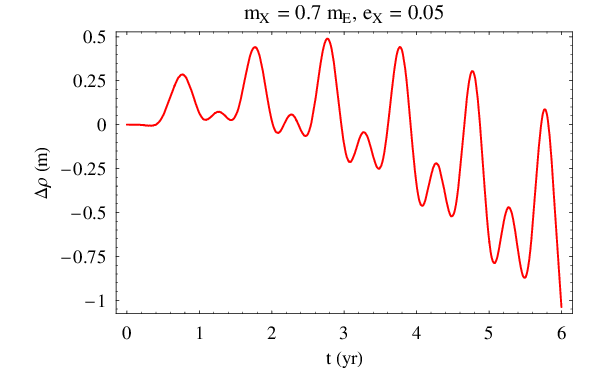,width=0.30\linewidth,clip=} & \epsfig{file=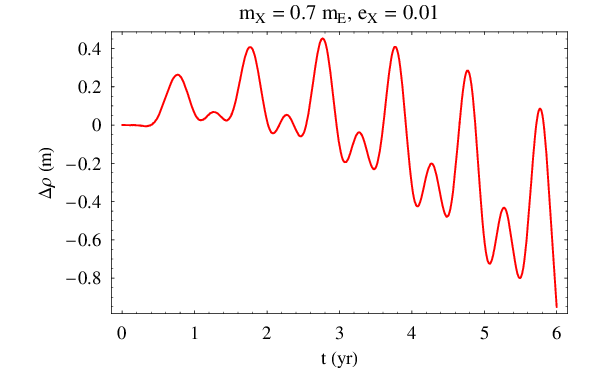,width=0.30\linewidth,clip=}\\
\end{tabular}
\caption{Numerically integrated Earth-New Horizons range signal $\Delta\rho$, in m, for $t_0= 1$ Jan 2015, $m_{\rm X}=0.7\ m_{\oplus}$, $a_{\rm X}=4,500$ au, $\Om_{\rm X}=60$ deg, $f_{\rm X}=0$ deg, $I_{\rm X}=120$ deg, $\omega_{\rm X}=50$ deg.}\lb{figura5}
\end{figure}
\begin{figure}
\centering
\begin{tabular}{ccc}
\epsfig{file=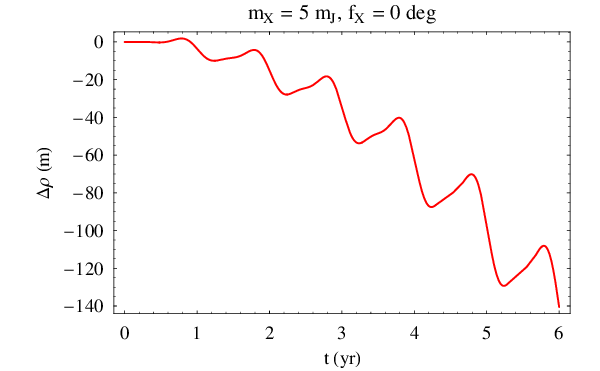,width=0.30\linewidth,clip=} & \epsfig{file=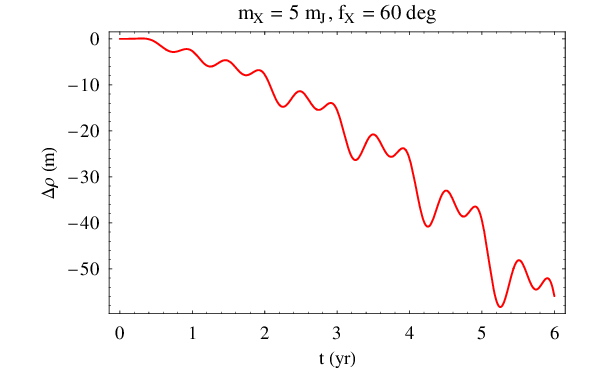,width=0.30\linewidth,clip=}&
\epsfig{file=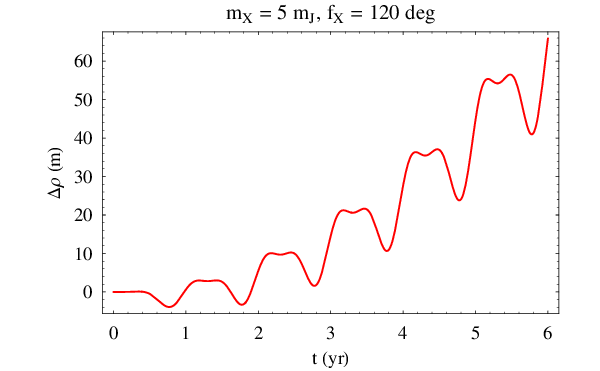,width=0.30\linewidth,clip=} \\
\epsfig{file=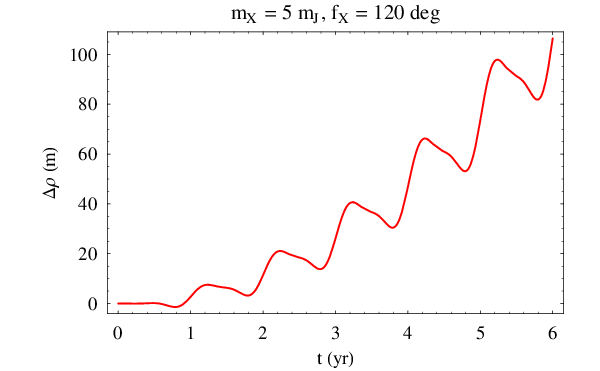,width=0.30\linewidth,clip=}&
\epsfig{file=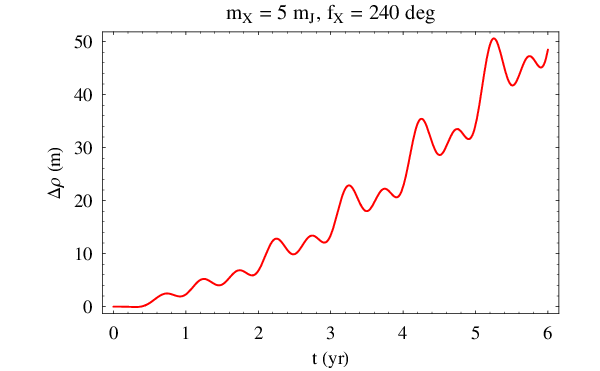,width=0.30\linewidth,clip=} & \epsfig{file=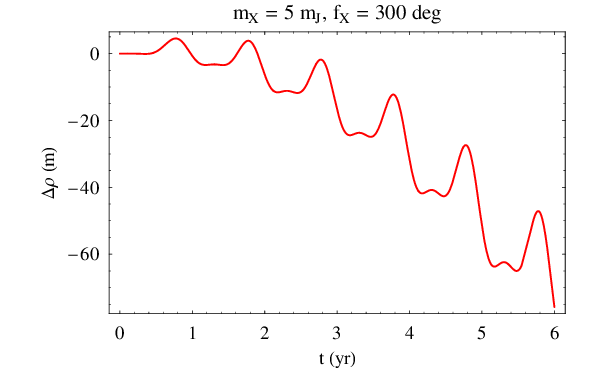,width=0.30\linewidth,clip=}\\
\end{tabular}
\caption{Numerically integrated Earth-New Horizons range signal $\Delta\rho$, in m, for $t_0= 1$ Jan 2015, $m_{\rm X}=5\ m_{\rm J}$, $a_{\rm X}=60,000$ au, $e_{\rm X}=0.07$, $I_{\rm X}=35$ deg, $\Om_{\rm X}=60$ deg, $\omega_{\rm X}=50$ deg.}\lb{figura6}
\end{figure}
\begin{figure}
\centering
\begin{tabular}{ccc}
\epsfig{file=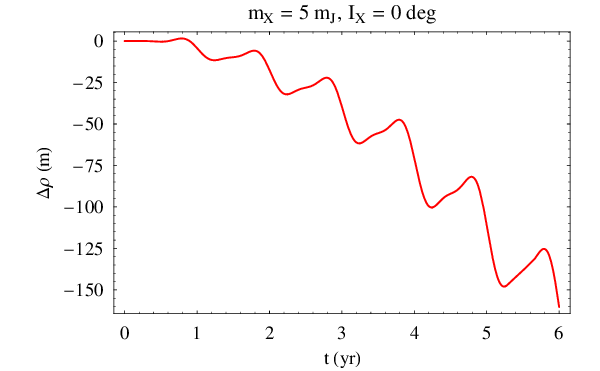,width=0.30\linewidth,clip=} & \epsfig{file=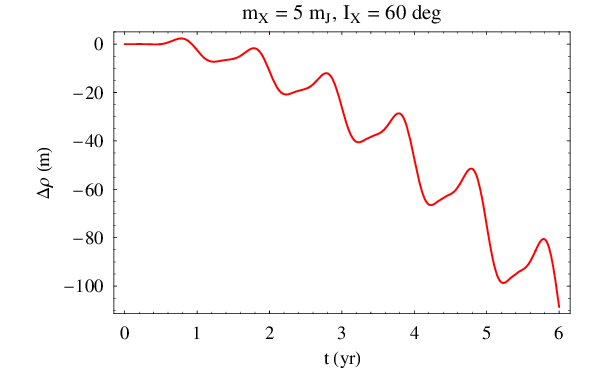,width=0.30\linewidth,clip=} &
\epsfig{file=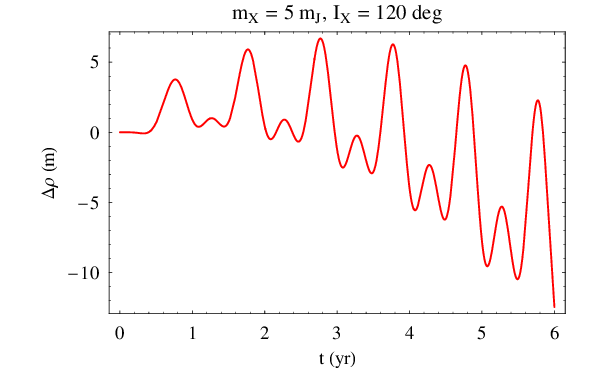,width=0.30\linewidth,clip=} \\
\epsfig{file=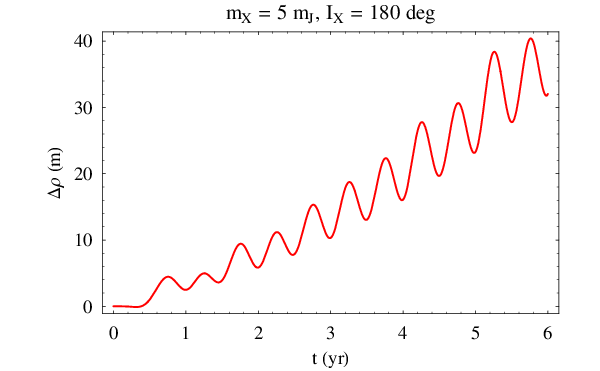,width=0.30\linewidth,clip=}&
\epsfig{file=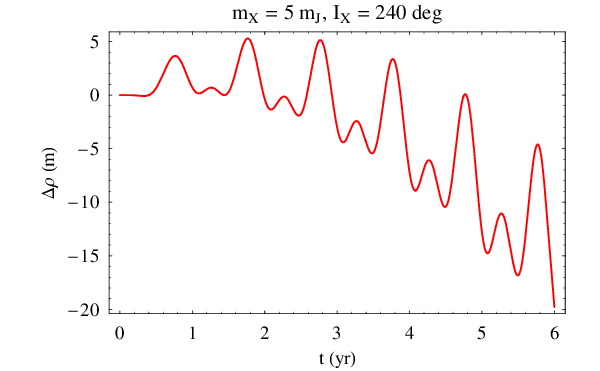,width=0.30\linewidth,clip=} & \epsfig{file=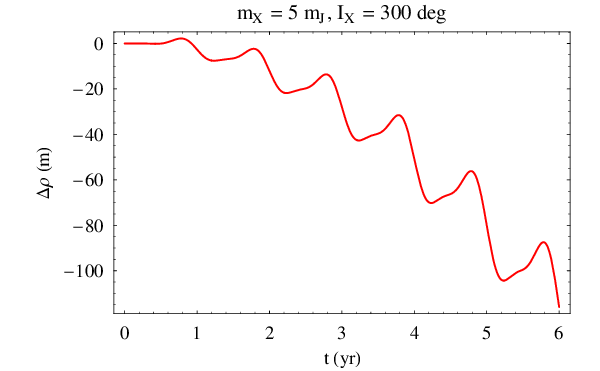,width=0.30\linewidth,clip=}\\
\end{tabular}
\caption{Numerically integrated Earth-New Horizons range signal $\Delta\rho$, in m, for $t_0= 1$ Jan 2015, $m_{\rm X}=5\ m_{\rm J}$, $a_{\rm X}=60,000$ au, $e_{\rm X}=0.07$, $f_{\rm X}=0$ deg, $\Om_{\rm X}=60$ deg, $\omega_{\rm X}=50$ deg.}\lb{figura7}
\end{figure}
\begin{figure}
\centering
\begin{tabular}{ccc}
\epsfig{file=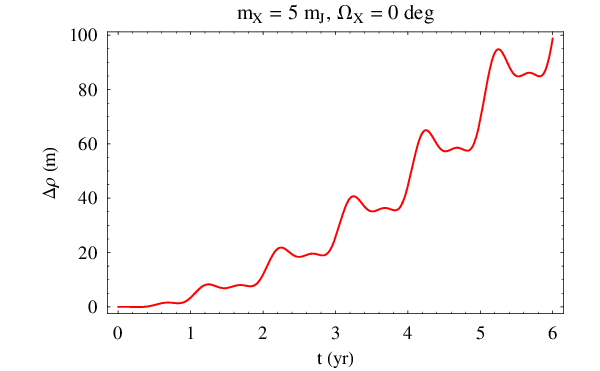,width=0.30\linewidth,clip=} & \epsfig{file=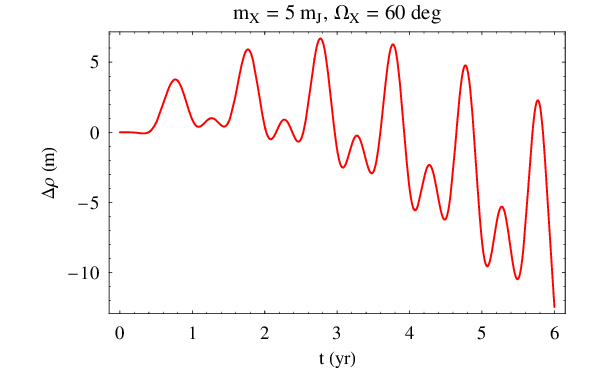,width=0.30\linewidth,clip=} &
\epsfig{file=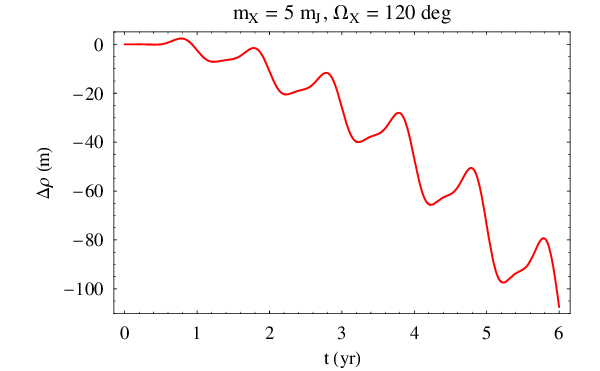,width=0.30\linewidth,clip=} \\ \epsfig{file=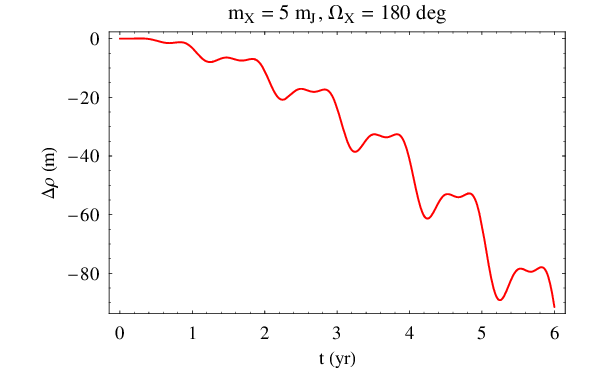,width=0.30\linewidth,clip=}&
\epsfig{file=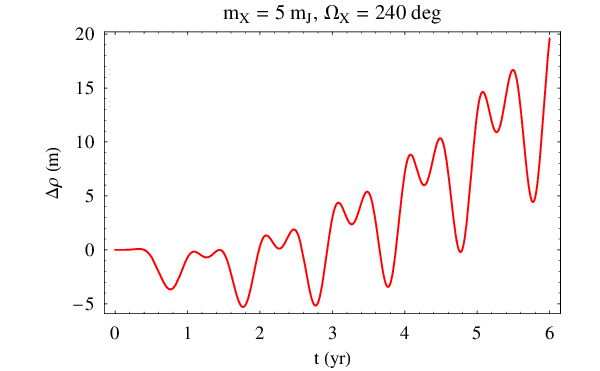,width=0.30\linewidth,clip=} & \epsfig{file=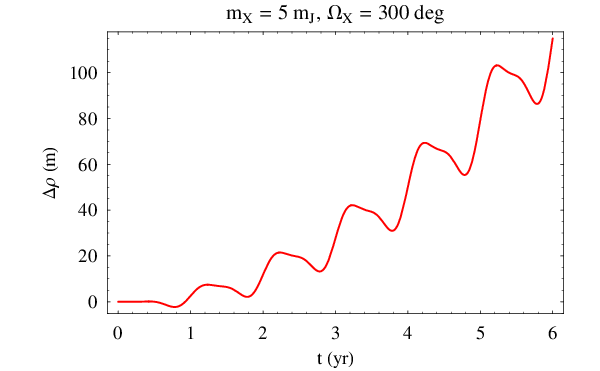,width=0.30\linewidth,clip=}\\
\end{tabular}
\caption{Numerically integrated Earth-New Horizons range signal $\Delta\rho$, in m, for $t_0= 1$ Jan 2015, $m_{\rm X}=5\ m_{\rm J}$, $a_{\rm X}=60,000$ au, $e_{\rm X}=0.07$, $f_{\rm X}=0$ deg, $I_{\rm X}=120$ deg, $\omega_{\rm X}=50$ deg.}\lb{figura8}
\end{figure}
\begin{figure}
\centering
\begin{tabular}{ccc}
\epsfig{file=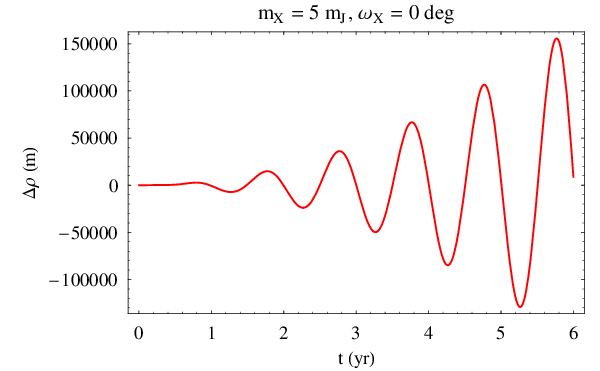,width=0.30\linewidth,clip=} & \epsfig{file=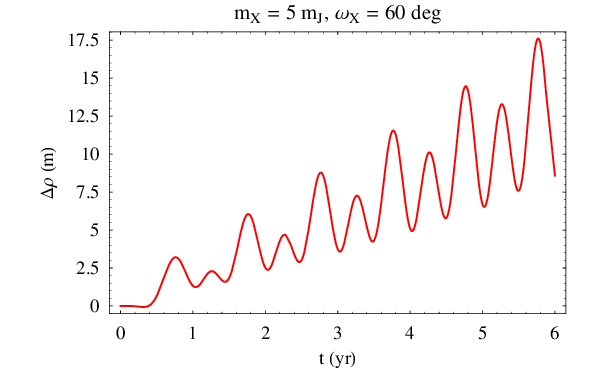,width=0.30\linewidth,clip=}&
\epsfig{file=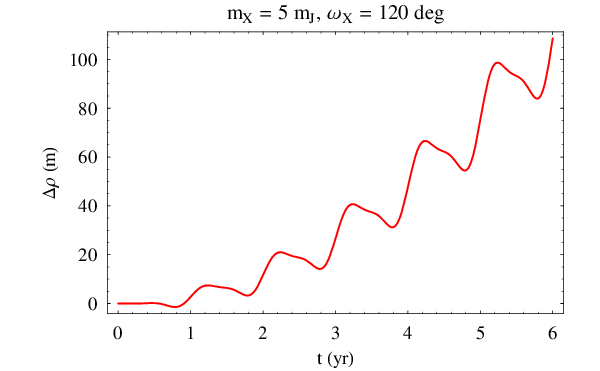,width=0.30\linewidth,clip=} \\
\epsfig{file=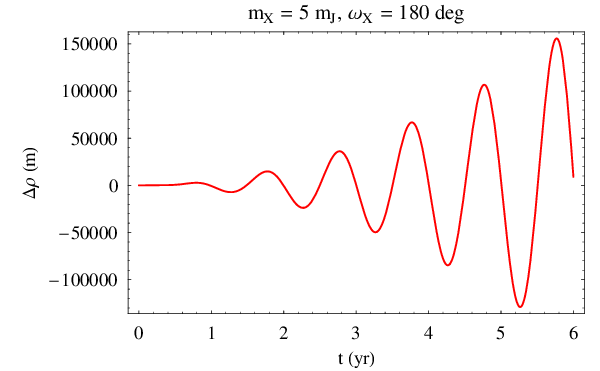,width=0.30\linewidth,clip=}&
\epsfig{file=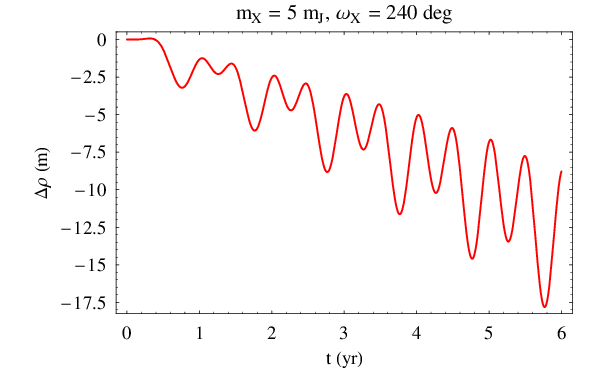,width=0.30\linewidth,clip=} & \epsfig{file=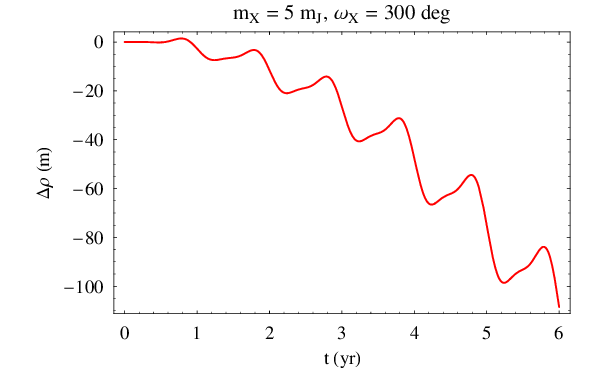,width=0.30\linewidth,clip=}\\
\end{tabular}
\caption{Numerically integrated Earth-New Horizons range signal $\Delta\rho$, in m, for $t_0= 1$ Jan 2015, $m_{\rm X}=5\ m_{\rm J}$, $a_{\rm X}=60,000$ au, $e_{\rm X}=0.07$, $f_{\rm X}=0$ deg, $I_{\rm X}=120$ deg, $\Om_{\rm X}=60$ deg.}\lb{figura9}
\end{figure}
\begin{figure}
\centering
\begin{tabular}{ccc}
\epsfig{file=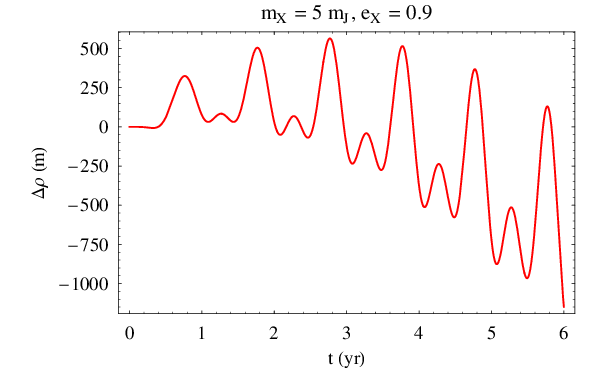,width=0.30\linewidth,clip=} & \epsfig{file=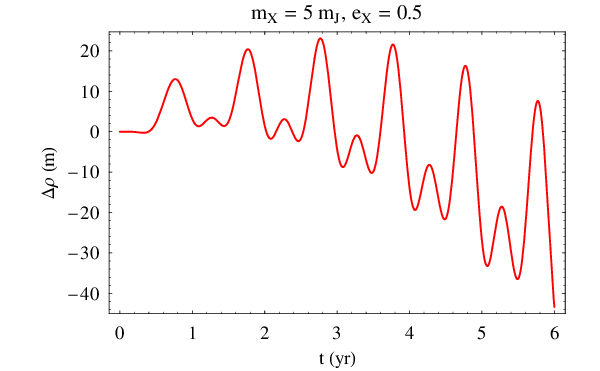,width=0.30\linewidth,clip=}&
\epsfig{file=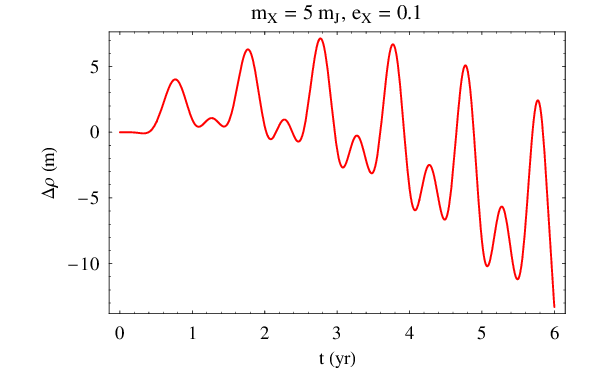,width=0.30\linewidth,clip=} \\
\epsfig{file=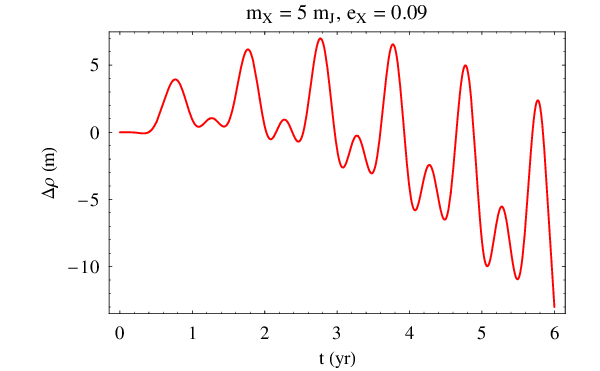,width=0.30\linewidth,clip=}&
\epsfig{file=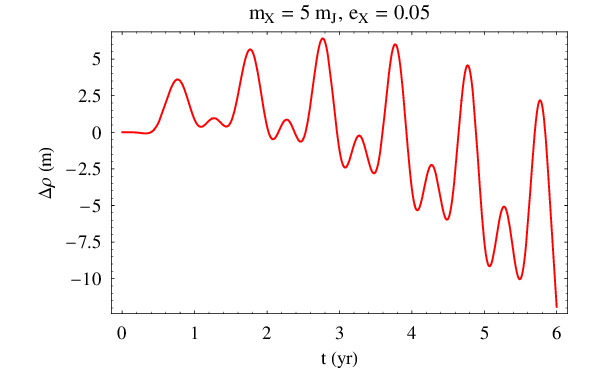,width=0.30\linewidth,clip=} & \epsfig{file=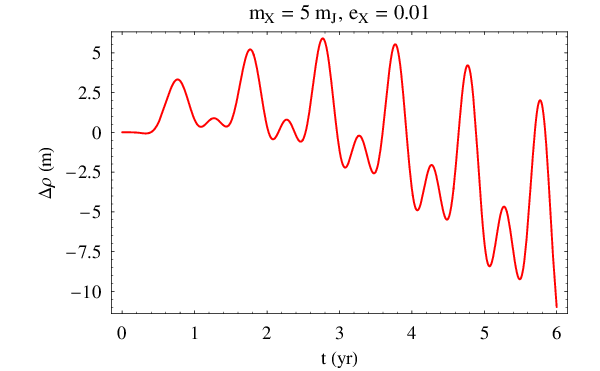,width=0.30\linewidth,clip=}\\
\end{tabular}
\caption{Numerically integrated Earth-New Horizons range signal $\Delta\rho$, in m, for $t_0= 1$ Jan 2015, $m_{\rm X}=5\ m_{\rm J}$, $a_{\rm X}=60,000$ au, $\Om_{\rm X}=60$ deg, $f_{\rm X}=0$ deg, $I_{\rm X}=120$ deg, $\omega_{\rm X}=50$ deg.}\lb{figura10}
\end{figure}

\bibliography{NHbib,SEPbib,QUMONDbib}{}
\end{document}